\documentstyle[12pt,epsf]{article}
\begin{document}
\def \cl {{\cal L}}
\def \d {{\rm d}}
\def \be {\begin{equation}}
\def \ee {\end{equation}}
\def \bea {\begin{eqnarray}}
\def \eea {\end{eqnarray}}
\def \bi {\bibitem}
\def \ci {\cite}
\def \e {{\rm e}}
\def \o {\omega}
\def \a {\alpha}
\def \n {\nu}
\def \cf {{\cal F}}
\def \x {\xi}
\def \g {\gamma}
\def \del {\partial}
\def \pt {{_{\rm PT}}}
\def \eps {\varepsilon}
\newcommand\re[1]{(\ref{#1})}
\def \lab #1 {\label{#1}}

\def \p {\pi}
\def \m {\mu}
\def \cs {{\cal S}}
\def \as {{\alpha_s}}
\def \CO {{\cal O}}

\newcommand \vev [1] {\langle{#1}\rangle}

\def\thefootnote{\fnsymbol{footnote}}
\thispagestyle{empty}

\begin{flushright}
\begin{tabular}{l}
ITP--SB--98--73 \\
LPT--Orsay--98--80 \\
hep-ph/9902341
\end{tabular}
\end{flushright}

\vspace{30mm}
\begin{center}
{\Large \bf Power Corrections to Event Shapes\\[3mm]
and Factorization}
\end{center}
\vspace{2mm}
\begin{center}
{\large Gregory P.\ Korchemsky}\\
\vspace{2mm}
{\it Laboratoire de Physique Th\'eorique
\footnote{Laboratoire associ\'e au Centre National de la Recherche
Scientifique (UMR 8627)}
\\
Universit\'e de Paris XI, Centre d'Orsay, b\^at 210 \\
91405 Orsay C\'edex, France} \\
\vspace{4mm}
{\large George Sterman}\\
\vspace{2mm}
{\it Institute for Theoretical Physics\\
State University of New York at Stony Brook\\
Stony Brook, NY 11794-3840, USA} \\
\vspace{2mm}
\end{center}

\begin{abstract}
We study power corrections to the differential thrust, heavy mass and
related event shape distributions in ${\rm e}^+{\rm e}^-$ annihilation,
whose values, $e$, are proportional to jet masses in the two-jet limit,
$e\to 0$. The factorization properties of these
differential distributions imply that
they may be written as convolutions of  nonperturbative ``shape" functions,
describing the emission of soft quanta by the jets, and
resummed perturbative cross sections. The infrared shape functions are
different for different event shapes, and depend on a
factorization scale, but
are independent of the center-of-mass energy $Q$. They organize all power
corrections of the form $1/(eQ)^n$, for arbitrary $n$,
and carry information on a class of universal matrix elements of the
energy-momentum tensor in QCD, directly related to the energy-energy
correlations.
\end{abstract}

\newpage

\def\thefootnote{\arabic{footnote}}
\setcounter{footnote} 0

\section{Introduction}

Infrared safe event shapes in ${\rm e}^+{\rm e}^-$ annihilation are finite
order-by-order in perturbation theory and can be calculated at large
center-of-mass energy $Q$, in powers of $\as(Q)$. Nonperturbative
corrections decrease as powers of $Q$ relative to the perturbative
series. In most cases such corrections are readily observed in
the data up to the highest available energies. They can be parameterized
as $\lambda_p/Q^p$, with scales $\lambda_p$ and exponents $p$ depending on the
shape in question, and they must be taken into account in
precise measurements of $\as(Q)$ based on event shapes
\ci{shapedata1,shapedata2}.

Some time ago, it was shown that perturbation theory itself suggests the
exponents $p$ (but not the scales $\lambda_p$)
for the leading power corrections to the mean values of
various event shapes \cite{W,KS1,AZ,BB}. The simplicity of these
results, and the successes of fits to the data based upon them, were
somewhat surprising, because previously nonperturbative effects were
estimated primarily by comparison to event generators.

Although perturbation theory cannot predict the magnitudes of the
scales $\lambda_p$, it is natural to entertain the possibility that the
leading power corrections to event shapes  might be controlled by a set of
nonperturbative but universal parameters \cite{KS1,AZ,KS2}.
The latter have been parameterized as moments of a universal effective
coupling over low momentum scales \ci{DMW}, and
further analyses incorporate effects of the effective
coupling beyond lowest order \ci{Milano}.
It is not clear, however, that long-distance effects
can be expressed fully in terms of low powers
of the running coupling, extrapolated to scales of order $\Lambda_{\rm QCD}$.
We therefore believe that a more general viewpoint will be desirable,
to further reveal the underlying QCD dynamics.

In this paper, we suggest a formalism for power corrections to IR safe
event shapes, based on the factorization of short-distance perturbative and
long-distance nonperturbative effects. The factorization formalism can be
thought of as a generalization of
the operator product expansion (OPE), applicable to weighted cross sections.
For event shapes, we encounter a new set of nonperturbative
distributions of soft radiation, which we shall term ``infrared shape
functions"
\ci{Moriond98}.
As we observe below, our infrared shape functions generalize
the inclusive (light-cone)
distributions that have
been introduced to describe semileptonic $b$ quark decays
near the edge of phase space \ci{KGb}.  They
admit operator definitions, and are related to
correlation functions of the energy-momentum tensor, which have been discussed
in a number of related contexts \ci{ST,KOS,T}. The moments of the
infrared shape functions
define the nonperturbative scales $\lambda_p$
that determine the mean values of
various event shapes. Previous descriptions of power corrections
reemerge as particular choices for these functions.

Rather than analyzing only mean values, we will treat {\it differential\/}
event shape distributions. The structure of power corrections in this case
is more complex, and more interesting.
For example, it has been shown for the thrust that when $t\equiv 1-T$ is
small, but substantially larger than $\Lambda_{\rm QCD}/Q$, the $1/Q$
power corrections produce a shift in the
resummed perturbative distribution \ci{KS1,KS2,DW},
\be
{d\sigma_\pt(t) \over
dt} \to {d\sigma_\pt(t-\lambda_1/Q) \over dt}
+\CO(1/(tQ)^2)\, ,
\lab{shift}
\ee
with $\lambda_1$ the same scale that parameterizes the $1/Q$
correction to
the mean thrust, $\vev{t}\sim {\lambda_1}/{Q}$.  This approach improves the
fit of the resummed perturbative formula for low values of $t$ and related
event shapes.  At the same time, Eq.\ (\ref{shift}) must break down
for very small values of $t\sim \Lambda_{\rm QCD}/Q$, corresponding to
small invariant mass jets in the final state, because in this region
all power corrections of the form $1/(tQ)^a$, with $a=1,2,\dots$
become equally important.
We shall observe that the factorization properties of
event shape distributions enable us to organize all
powers of $1/(tQ)$, by expressing the distributions
as convolutions of nonperturbative infrared shape functions with resummed
perturbative spectra.
The scale for this factorization, $\mu \gg \Lambda_{\rm QCD}$,
will act as an infrared (IR) cutoff in the perturbative expressions.
Although the physical distributions are independent of $\mu$, the shape
function and the resummed spectrum for each event shape depend, as usual,
on the factorization scale. The resulting expressions for the
differential distributions are applicable over a wide interval
of the event shapes, and for moderate $t$ they take the simple form
of Eq.\ \re{shift}.

We shall discuss two representative event shapes in some detail,
the thrust, $t$, and the heavy jet mass distribution, $\rho$.
These two shapes are ideal testing-grounds for our proposals on the
interplay of perturbative and nonperturbative effects in QCD,
since we have fairly good control over their perturbative
expansions, while at the same time the data indicate the presence of
significant power corrections.

We begin, in the following section, by showing that the
kinematics of ${\rm e}^+{\rm e}^-$ annihilation near
the two-jet limit suggests that the heavy jet
mass and thrust distributions are convolutions of
nonperturbative and perturbative components.
In Sec.\ 3 we show that the same convolution
form emerges from known factorization properties of the resummed
cross sections in the two-jet limit.
We go on in Sec.\ 4 to show how
the infrared shape functions are
related to the expectation values of products
of Wilson lines, independent of the center-of-mass energy.
We then show how the truely universal nonperturbative
information in event shapes is encoded in a set of correlation functions
of an energy flow operator, related to energy-energy correlations.

\section{The Radiation Function and Factorization}

Consider an event shape $e\ (=1-T\,,\rho\,,\dots)$, which takes on a value
$e(N)$ for each ${\rm e}^+{\rm e}^-$ annihilation final state $|N\rangle$.
We represent the differential distribution $d\sigma/de$ schematically by
\be
{1\over \sigma_{\rm tot}}\; {d\sigma(Q)\over de}
= {1\over \sigma_{\rm tot}}\;
\sum_N\, \left|\left\langle N\left|\; J(0)\; \right|0\right\rangle\right |^2
\, \delta\left(e-e(N)\right)\equiv \left\langle
\delta\left(e-e(N)\right)\right\rangle\,,
\lab{dsigdef}
\ee
where $J(x)$ represents the appropriately normalized electroweak current,
and the sum over states $N$ is taken at fixed momentum $q$, $q^2=Q^2$.
The second equality in
Eq.\ (\ref{dsigdef}) defines
a convenient notation for the weighted distribution.

The weights $e(N)$ for the thrust and jet mass
are given in the end-point region, $e\to 0$, in terms of the total invariant
masses $M_{R,L}^2$ flowing into right and left hemispheres defined
by the plane orthogonal to the thrust axis, as
\bea
t(N) &=& {M_R^2+M_L^2\over Q^2}\nonumber\\
\rho(N) &=& {M_R^2\over Q^2}\; \theta(M_R^2-M_L^2)
+ {M_L^2\over Q^2}\; \theta(M_L^2-M_R^2)\, .
\lab{trhodef}
\eea
Each soft particle with momentum $\ell_k=(E_k,\vec \ell_k)$
in the final state contributes to these weights
additively. Its contribution depends on whether it goes into the
right $(\cos\theta_k>0)$ or left $(\cos\theta_k<0)$ hemisphere, with
$\theta_k$ the angle between the 3-momentum of the particle, $\vec \ell_k$,
and the
thrust axis. Neglecting terms quadratic in the energy of the soft
particles, we find
\be
M^2_{R,L}=Q
\sum_{k=1}^{N_{R,L}}\; E_k(1-|\cos\theta_k|)\, ,
\lab{m2rl}
\ee
where $N_{R,L}$ denotes the number of soft particles in the corresponding
hemisphere.

We now introduce a ``radiation function", $R(e)$,
in the notation of Eq.\ (\ref{dsigdef}), through \ci{CTTW}
\be
R(e) = \left\langle \theta\left(e-e(N)\right) \right\rangle
\equiv
\int_0^{e} de'\, {1\over \sigma_{\rm tot}}\; {d\sigma\over de'}
=1-\int_{e}^{e_{\rm max}}de'\, {1\over \sigma_{\rm tot}}\; {d\sigma\over
de'}\, .
\lab{Redef}
\ee
Here, $e_{\rm max}$ is the maximum value of the event shape, $e$, and 
$\sigma_{\rm tot}$ is the integral of
the differential distribution over its full range,
$\sigma_{\rm tot}=\int_0^{e_{\rm max}}de\, d\sigma(e)/de$.
The function $R(e)$ is unity at $e=e_{\rm max}$ and vanishes
as $e\to 0$.  We expect perturbation theory to be accurate for $R(e)$ at
``large" $e\sim e_{\rm max}$, and to fail when $e={\cal O}(\Lambda/Q)$.
In the notation of Eq.\ \re{dsigdef}, the radiation function for the heavy
jet mass is
\be
R_H(\rho)
= \left\langle \theta\left(\rho-\rho(N)\right) \right\rangle
= \left\langle \theta\left(\rho-{M_R^2\over Q^2}\right)\,
\theta\left(\rho-{M_L^2\over Q^2}\right)\right\rangle\, ,
\lab{RH}
\ee
with $M_R$ and $M_L$ given in Eq.\ (\ref{m2rl}).
Similarly, the radiation function for the thrust is
\be
R_T(t) = \left\langle \theta\left(t-t(N)\right) \right\rangle
= \left\langle \theta\left(t-{M_R^2+M_L^2\over
Q^2}\right)\right\rangle\, .
\lab{RT}
\ee
The functions $R_H$ and $R_T$ receive both perturbative and
nonperturbative contributions.

For such event shapes, $e=0$ corresponds to a final state with two
infinitely narrow jets,
and the resummed perturbative
differential cross section vanishes in this limit,
while its expansion in $\alpha_s(Q)$ is singular order by order \cite{CTTW}.
For small $e$, the final state
consists of two narrow quark jets of invariant mass $\sim Qe^{1/2}$ and
a ``cloud" of soft particles of the total energy $\sim Qe$.
We expect nonperturbative corrections to occur
with inverse powers of both scales. However, as $e$ decreases, the energy
of the soft radiation reaches a nonperturbative scale before the jet
invariant masses, because $Qe^{1/2} \gg Qe$. Therefore, for $e^{1/2}\gg
\Lambda_{\rm QCD}/Q
\sim e$, we may restrict ourselves to power corrections in the smaller scale
$Qe$.    We shall see below that it is possible to organize systematically
power corrections of the
form $1/(Qe)^a$, with $a=1,2\dots$. With this in mind, we consider the limit
$e\to 0$ with $Qe=\rm fixed$, and neglect corrections suppressed by additional
inverse powers of $Q$.

In this limit, the jet invariant masses $M_{R,L}^2$ are
sensitive only to the light-cone components of
soft particle momenta in the final state,
$\ell^\pm_k=E_k(1\pm\cos\theta_k)/\sqrt{2}$, defined relative
to the thrust axis.
Assuming that nonperturbative effects are small,
we may separate the contributions of soft particle emission
to $M_L$ and $M_R$ as
\be
M_R^2 \to  M_R^2+\eps_R Q\,,\qquad
M_L^2 \to M_L^2+\eps_L Q\,,\qquad
\eps_{R,L} = \sum_{k=1}^{N^{\rm soft}_{R,L}} E_k(1-|\cos\theta_k|)\, ,
\lab{MtoMeps}
\ee
where from now on, $M_{R,L}^2$ refer to the perturbative contributions
to the hemisphere masses squared,
and $\eps_{R,L}Q$ to the soft-parton, nonperturbative
contributions. Because, as we now argue,
soft particle emission may be factorized from the jets,
$\eps_{L,R}$ are sums over the light-cone momentum components
of $N^{\rm soft}_{L,R}$ soft particles,
emitted into the respective hemispheres.

The basis of factorization is that the emission
of soft gluons occurs over time
scales that are different from those involved in
the evolution of narrow jets.  These processes
are quantum-mechanically incoherent, and
the soft gluon distribution
emitted by a pair of narrow jets at wide angles
depends only on the direction and total color
charge of the jets.  (See Sec.\ 4 below.)
Then, to all orders in perturbation theory,
inclusive cross sections for two narrow jets in $\e^+\e^-$ annihilation
can be written as products of separate functions for the
jets and  for the soft radiation,
convoluted in the light-cone components of
the soft radiation.  Corrections are
suppressed by powers of $M_{L,R}^2/Q^2$.
Details of the necessary reasoning are reviewed in Ref.\ \cite{factref}.
Here, we apply this factorization
to the limit $e\to 0$, $Qe$ fixed, identified above.
In this limit, the jets
are narrow, but their invariant masses are still large
enough for perturbation  theory to be valid, while the
soft-gluon function  becomes nonperturbative.

For the heavy jet mass, we use Eqs.\
\re{RH} and \re{MtoMeps} to identify the appropriate factorized
expression as
\be
R_H(\rho) = \int_0^\mu d\eps_R \int_0^\mu d\eps_L\; f(\eps_R,\eps_L)\,
\left\langle
\theta\left(\rho-{M_R^2\over Q^2}-{\eps_R\over Q}\right)\,
\theta\left(\rho-{M_L^2\over Q^2}-{\eps_L\over Q}\right)\right\rangle_\pt\, ,
\lab{RHp}
\ee
where the subscript PT denotes the average with respect to the
perturbative spectrum in Eq.\ (\ref{Redef}),
in a  manner we will specify below.
The factor $f(\eps_R,\eps_L)$ is the nonperturbative infrared shape function
referred to above.  It represents the probability density
for the total
soft gluon light cone momentum components,
$\eps_R$ and $\eps_L$,
in each hemisphere.  The factorization of soft
dynamics implies that $f(\eps_R,\eps_L)$ does not depend on the hard scale
$Q$, up to corrections suppressed by $1/Q$,
although it does depend on the cut-off $\mu$ that sets  the
maximal energy of particles described by the shape function.
The $\mu-$dependence of the shape function is compensated in \re{RHp}
by that of the perturbative contribution. (For now, we suppress $\m$ as an
argument
for simplicity.)

For the thrust, we have, with the {\it same\/} shape function
$f(\eps_R,\eps_L)$,
\be
R_T(t) = \int_0^\mu d\eps_R \int_0^\mu d\eps_L\; f(\eps_R,\eps_L)\,
\left\langle
\theta\left(t-{M_R^2+M_L^2\over Q^2}-
{\eps_R+\eps_L\over Q}\right)\right\rangle_\pt\, .
\lab{RHps}
\ee
The shape functions in Eqs.\ \re{RHp} and \re{RHps} are identical precisely
because of
the factorization of soft gluon emission from the jets.
Note that $f(\eps_R,\eps_L)$ is a symmetric function of the
soft radiation variables $\eps_{R,L}$ separately. Because $R_T$
depends only on the sum of soft momenta, we may simplify
it by introducing the function
\be
f_T(\eps)=\int_0^\mu d\eps_R\int_0^\mu
d\eps_L\; f(\eps_R,\eps_L)\; \delta(\eps-\eps_R-\eps_L)\, ,
\lab{FTdef}
\ee
in terms of which
\be
R_T(t) = \int_0^\mu d\eps\; f_T(\eps)\; \left\langle
\theta\left(t-{\eps\over Q}-{M_R^2+M_L^2\over
Q^2}\right)\right\rangle_{\pt}\, .
\lab{RTrev}
\ee
Thus, the leading power corrections to the thrust give rise to an
${\eps/Q}-$shift
of the perturbative radiation function, just as in Eq.\ \re{shift},
but now averaged with
the shape function $f_T(\eps)$.
In contrast, the jet mass function, Eq.\ \re{RHp}, does
not have this property, unless
\be
f(\eps_L,\eps_R)=f(\eps_R)\; f(\eps_L)+\Delta f(\eps_L,\eps_R)
\, ,
\lab{ffact}
\ee
with $\Delta f$ small,
corresponding to independent evolution for the two hemispheres.
This is not obvious even
perturbatively, however, and
$\Delta f(\eps_L,\eps_R)$ is nonzero, for example, due to the decay of
an off-shell gluon into a pair of gluons, each moving into different
hemispheres \ci{NS}.

Even though Eq.\ \re{ffact} is not a general property, the perturbative
expectation values
 do factorize into separate functions for the jets
to resummed next-to-leading logarithmic approximation (NLL) \ci{CTTW},
\be
\left\langle
\theta\left(\rho'-{M_R^2\over Q^2}\right)\,
\theta\left(\rho'{}'-{M_L^2\over Q^2}\right)\right\rangle_\pt
\stackrel{{\rm NLL}}{=}
\left\langle
\theta\left(\rho'-{M_R^2\over Q^2}\right)
\right\rangle_\pt
\left\langle
\theta\left(\rho'{}'-{M_L^2\over Q^2}\right)\right\rangle_\pt\, .
\lab{thetatheta}
\ee
We will introduce the infrared cutoff into these perturbative
functions.  Following Eq.\ \re{thetatheta}, we for the
heavy jet mass distribution,
\be
R_H(\rho) = \int_0^{\rho Q}d\eps_R\int_0^{\rho Q}d\eps_L\,
f(\eps_R,\eps_L,\m)\, R_J^\pt\left(\rho-{\eps_R\over Q},\m\right)\;
R_J^\pt\left(\rho-{\eps_L\over Q},\m\right)\, ,
\lab{RHRjet}
\ee
where  $R_J^\pt(\rho,\mu)$ is the cutoff-dependent perturbative
radiation function for
a single jet with invariant mass $\rho Q^2$,
\be
R_J^\pt(\rho,\mu) =
\left\langle \theta \left(\rho-{M^2\over Q^2} \right)\right\rangle_{\rm PT}
=
\theta(\rho)
\left[1 -
\int_{{\rm max}(\rho,\mu/Q)}^{\rho_{\rm max}} d\rho'\;
{1\over \sigma_{\rm tot}}\, {d\sigma_J^{\rm PT}\over d\rho'}
\right]
\, .
\lab{Rjetdef}
\ee
In this expression, ${d\sigma_J^\pt/d\rho'}$
is the inclusive perturbative cross section
for the creation of a single jet of invariant mass $\rho Q^2$, and
we have exhibited the $\m$ dependence for clarity.
Compared with Eq.\ \re{Redef}, an infrared cutoff is imposed on $\rho$,
by demanding that $\rho Q^2\ge \mu Q$, which corresponds to the
mass of a jet that includes radiation with energy
of order $\m$ at large angles to the
jet axis.  In this way, we introduce $\m$ as the factorization scale
that separates perturbative and nonperturbative
dynamics. The perturbative radiation function,
$R^\pt_J$, is ``frozen" for $0<\rho<\mu/Q$. This is
equivalent to cutting off the perturbative spectrum at the value $\rho=\mu/Q$.
All dynamics below $\rho=\mu/Q$ is to be incorporated into the soft shape
function.
Other choices for the transition between perturbative and nonperturbative
regions are possible, but this is relatively simple to implement
numerically \cite{Moriond98}.
Note that $R_J^\pt(0,\m)=R_J^\pt(\mu/Q,\m)$
includes the contributions only of virtual
soft gluons with energies above $\mu$.
Therefore, because $\mu$ is nonzero,
infrared divergences do not appear and
$R_J^\pt(0,\m)$ does not vanish, although it is small.

For the thrust, the corresponding expression for $R_T(t)$ is
\be
R_T(t)=\int_0^{tQ}d\eps\; f_T(\eps)\; R_T^{\rm
PT}\left(t-{\eps\over Q},\mu\right)\, ,
\lab{RTRjet}
\ee
where $R_T^\pt(t',\mu)$ is the perturbative radiation function
corresponding to independent evolution of two jets, the sum of whose
invariant masses is bounded by $t'Q^2$:
\be
R_T^\pt(t',\mu)=\theta(t)\left[
1-
\int_{\mu/Q}^{t_{\rm max}} d\rho'
{1\over \sigma_{\rm tot}}\, {d\sigma_J^\pt\over d\rho'}
\int_{\mu/Q}^{t_{\rm max}} d\rho''
{1\over \sigma_{\rm tot}}\, {d\sigma_J^\pt\over d\rho''}\,
\theta\left(\rho'+\rho''-{\rm max}
\left\{t',\frac{\mu}{Q}\right\}
\right)
\right]\,.
\lab{R2jpt}
\ee
Here, the IR cut-off is implemented on the invariant mass of each jet and
on their sum.  In the remainder of this section, for simplicity we shall
again suppress
the argument $\m$ in the cut off perturbative and infrared shape
functions, although they are always present.

The thrust and heavy jet mass
distributions themselves
are found by taking derivatives of their corresponding radiation functions,
Eqs.\ \re{RTRjet} and \re{RHRjet}, respectively.
Consider first the thrust, for which
\be
{1\over \sigma_{\rm tot}}{d\sigma\over dt}
=
{dR_T\over dt}=Qf_T(tQ)\; R_T^\pt(0)
+\int_0^{t Q}d\eps\; f_T(\eps)\;
\frac1{\sigma_{\rm tot}^\pt}{d\sigma^\pt
\left(t-{\eps\over Q}\right)
\over dt}\, ,
\lab{dsigdt}
\ee
where
\be
\frac1{\sigma_{\rm tot}^\pt}{d\sigma^\pt\left(t'\right)\over dt}
=
{d R_T^\pt\left(t'\right)\over dt}\, .
\lab{dsigdRjet}
\ee
Here, the first term corresponds to freezing perturbative real soft gluon
radiation,
and replacing it with the nonperturbative shape function $f_T(tQ)$
(multiplied by the cut-off Sudakov form factor $R_T^\pt(0)$).
The second term is the perturbative spectrum smeared with the shape function.
The resulting form, Eq.\ (\ref{dsigdt}),
has a simple physical meaning. The
leading $1/(tQ)-$nonperturbative effects act independently
of perturbative branching, and the physical distribution is
obtained by convoluting the perturbative spectrum $d\sigma^\pt/dt$
with the nonperturbative infrared shape function $f_T(\eps)$.
Although $d\sigma^\pt/dt$ and
 $f_T(\eps)$ depend separately on the factorization scale $\mu$,
the physical cross-section $d\sigma/dt$ is $\mu-$independent.
As emphasized above, the nonperturbative shape function $f_T(\eps)$ depends on
$\mu$ but not on the hard scale $Q$. It takes its maximum value for
$\eps\sim\mu$ and rapidly vanishes for larger $\eps$. In contrast,
the perturbative spectrum starts at $t \sim \mu/Q$ and extends to
$t_{\rm max}$. For very small $0< t < \mu/Q$ the shape of
the thrust distribution is governed entirely by
$f_T(tQ)$.

For the heavy mass distribution we differentiate Eq.\ (\ref{RHRjet}) and
use the
symmetry of the shape function, $f(\eps_L,\eps_R)=f(\eps_R,\eps_L)$
to obtain
\be
{1\over \sigma_{\rm tot}}{d\sigma\over d\rho}
=Qf_H(\rho Q,\rho Q)R_J^\pt(0)+\int_0^{\rho Q}d\eps
\, f_H(\eps,\rho Q)\;\frac1{\sigma_{\rm tot}^\pt}{d\sigma_J^\pt
\left(\rho-{\eps\over Q}\right)
\over d\rho}\,,
\lab{dsigdrho}
\ee
where we have introduced the function $f_H$,
\be
f_H(\eps,\rho Q)=2\int_0^{\rho Q}d\eps'
\, f(\eps,\eps')\;
R_J^\pt\left(\rho-{\eps'\over Q}\right)\,.
\lab{fH}
\ee
Although the shape functions $f_T$ and $f_H$ in
Eqs.\ (\ref{dsigdt}) and (\ref{dsigdrho}) are related to the
same distribution $f(\eps_L,\eps_R)$, in general they are different.
In contrast with $f_T$, the shape function $f_H$ depends
both on the center-of-mass energy and the shape variable.
For $\rho Q >\mu$ we may expand the shape function (\ref{fH}) in powers
of $1/(\rho Q)$ as
\be
f_H(\eps,\rho Q)=2R_J^\pt(\rho)
\left[
\int_0^\mu d\eps' \, f(\eps,\eps')
-{1\over \rho Q}
\frac{d\ln R_J^\pt(\rho)}{d\ln \rho}
\int_0^\mu d\eps' \, \eps' f(\eps,\eps')
+\CO\left({1\over(\rho Q)^2}\right)
\right]\,.
\ee
Comparing Eqs.\ \re{dsigdt} and \re{dsigdrho}, we see that the leading
power corrections
to the thrust and the heavy mass distribution cannot
be described by a simple shift, associated, for instance, with a universal
effective coupling
constant.
It is, however,
relatively straightforward to use
Eqs.\ (\ref{dsigdt}) and \re{dsigdrho} phenomenologically
\cite{Moriond98}.

\section{Soft Radiation in Perturbation Theory}

Let us now show how the principal features of
nonperturbative shape functions emerge in resummed perturbation theory.
By deducing the $Q$-dependence of nonperturbative
contributions from ambiguities in the perturbative expansion
\ci{W,KS1,AZ,BB,KS2,DMW},
we will show that leading nonperturbative contributions
have the properties of the infrared shape functions
identified in Sec.\ 2.  Specifically, they enter the
cross section in convolution with
perturbative cross sections, and they are {\it independent of Q}.
To be specific, we will concentrate on the thrust distribution $d\sigma/dt$.

We start by recalling that the cross section $d\sigma/dt$
computed in the end-point region $t\sim 0$ by resumming an infinite number of
soft gluon emissions, exponentiates under a Laplace transform,
\be
{1\over \sigma_{\rm tot}}\,
\int_0^{t_{\rm max}} dt\, \e^{-\nu t} \frac{d\sigma}{dt} = \e^{-S(\nu,Q)}\, .
\lab{Laplace}
\ee
This property has been derived in a variety of ways
\cite{CTTW,gatheraletc,CLS}, and is related, in particular,
to the independent evolution of the opposite-moving jets.
The exponent $S(\nu,Q)$ in Eq.\ (\ref{Laplace}) is of the general form
\bea
S(\nu,Q) &=& \int_0^1 {d\alpha \over \alpha}\; \left( 1-\e^{-\nu\alpha}
\right)\;
\left [
\int_{\alpha^2Q^2}^{\alpha Q^2} {dk_\perp^2\over\ k_\perp^2}\;
\Gamma \left(\alpha_s(k_\perp^2)\right)
+
B\left(\as(\alpha Q^2)\right)\; \right]
\nonumber\\
&\equiv& S_{\rm PT}(\nu,Q,\mu)+S_{\rm NP}(\nu,Q,\mu)\,,
\lab{dec}
\eea
where in the absence of an IR cut-off the integration over soft gluon
momenta extends to arbitrarily small values of $k_\perp$.
As indicated in the second equality, we will separate perturbative and
nonperturbative
parts of $S$ by introducing
a factorization scale $\mu$. Heuristically, the contributions
of gluons with $k_\perp > \mu$ are to be absorbed into $S_\pt$, and those
with $k_\perp < \mu$ into a new, nonperturbative
exponent $S_{\rm NP}$.

Consider first the perturbative exponent $S_\pt(\nu,Q,\mu)$. Its properties
depend on the value of the Laplace transform parameter $\nu$, which in turn is
conjugate to the thrust variable, $1/\nu\sim t$. For $1/\nu>\mu/Q$ one
can replace $\left( 1-\e^{-\nu\alpha} \right)\rightarrow
\theta(\alpha-1/\nu\e^{-\gamma_E})$ in (\ref{dec}) and derive an expression
for $S_\pt$ that resums all leading and next-to-leading logarithms of
$\nu$ \cite{CTTW}.

For $\a>1/\nu$, the coupling in the term $B(\as(\alpha Q^2))$
is always at a perturbative scale, and $S_{\rm NP}$ comes entirely
from the $\Gamma$ term in Eq.\ \re{dec}.
In general, $S_{\rm NP}$ depends on two scales, $\nu/Q$ and $\nu/Q^2$.
Formally expanding in powers of $\nu/Q$, we find
\bea
S_{\rm NP}(Q/\nu,\mu)
&\equiv&
\int_0^1 {d\alpha \over \alpha}\; \left( 1-\e^{-\nu\alpha} \right)\;
\int_{\alpha^2Q^2}^{\alpha Q^2} {dk_\perp^2\over\ k_\perp^2}\;
\Gamma \left(\alpha_s(k_\perp^2)\right)
\nonumber\\
&=&
\sum_{n>0}\; {1\over n n!}\; \left( {-\nu\over Q}\right)^n\; \int_0^{\mu^2}\;
dk_\perp^2\ k_\perp^{n-2}\; \Gamma \left(\alpha_s(k_\perp^2)\right)
\nonumber\\
&\equiv&
\sum_{n>0}\; \frac1{n!}\left( {\nu\over Q}\right)^n\; \lambda_n(\mu)\, ,
\lab{Girexpland}
\eea
where we have neglected all powers of $\nu/Q^2$,
which are suppressed by powers of $1/Q$ relative to those we have kept.
In the final line of Eq.\ \re{Girexpland},
we absorb the ambiguous integrals over soft gluon momenta into
the definitions of nonperturbative
but $Q$-independent parameters $\lambda_n(\mu)$.
Since the variable $\nu$ is conjugate
to $t$, the final line of \re{Girexpland} organizes all power corrections in
$1/(tQ)$.
This infinite set of nonperturbative
parameters $\lambda_n(\mu)$, defines a single, $Q$-independent
function through
\be
\int_0^\infty d\eps \, \exp(-\nu\eps/Q) f_T(\eps;\mu)
=\exp\left(-S_{\rm NP}(Q/\nu,\mu)\right)\,,
\lab{f}
\ee
where, as the notation suggests, we may identify this function
with the infrared shape function for the thrust cross section.

Comparing the expansions of the two sides of this relation
in powers of $\nu/Q$ we derive the sum rules
\be
\int d\eps \, f_T(\eps;\mu) = 1\,,\quad
\int d\eps\,\eps\, f_T(\eps;\mu) = \lambda_1\,,\quad
\int d\eps\, \eps^2\, f_T(\eps;\mu) = \lambda_1^2-\lambda_2 \,, \dots
\lab{avgepsn}
\ee
Eqs.\ \re{dec} and \re{Girexpland} give
the Laplace transform of the thrust distribution \re{Laplace}
as a product of perturbative and nonperturbative functions
\footnote{Although this
result was obtained without use of the OPE the same property holds for
quantities which admit the OPE (for example, DIS structure functions for
$x\to 1$).}.
Replacing the nonperturbative exponent in Eq.\ \re{dec} by its expression
\re{f}
and performing the inverse Laplace transformation, we obtain
the thrust distribution $d\sigma/dt$ in the form of a convolution of the
perturbative Sudakov spectrum $d\sigma_\pt/dt$ and a $Q$-independent
nonperturbative shape function $f_T(\eps,\m)$, as in Eq.\ (\ref{dsigdt}).
 
We emphasize that this formalism is designed to treat $t\ge \mu/Q$, or
equivalently $\nu \le Q/\mu$.  Beyond this region, the soft function acquires
$Q$-dependence through terms that are neglected here,
including those associated with $B$ in Eq.\ \re{dec}.  Similarly,
for $\nu>Q/\mu$, it is necessary to go beyond the NLL approximation
in $S_\pt$.  We shall defer the treatment of these extentions
to future work.

Summarizing our analysis of soft gluon resummation,
we conclude that, in accordance with our expectations, the factorized
expression for the thrust distribution, Eq.\ \re{dsigdt}, resums large
perturbative logarithms and organizes all power corrections of the form
$(1/tQ)^n$.
The latter are described by a single infrared shape function $f_T(\eps)$
that takes
into account the contribution to the thrust of soft gluons with energy
smaller than $\mu$. It is straightforward to repeat similar analysis for
the heavy mass distribution to arrive at Eq.\ \re{dsigdrho}.

\section{Correlation Functions}

We now turn to the field-theoretic content of the shape functions.
Because the IR dynamics of the soft gluon radiation described by
$f(\eps_R,\eps_L)$ and $f_T(\eps)$ factorizes from the
quark jets, we can apply the eikonal approximation,
and replace the jets by eikonal lines in the computation of the soft emission.
Let $p_+$ and $p_-$ be the light cone directions defined by the quark jets.
The corresponding eikonal, or ``Wilson", lines, which serve as sources
of soft radiation, may be written as
$\Phi_\pm(0)= {\rm P}\exp \left[ig\int_0^\infty ds\; p_\pm^\mu
A_\mu(p_\pm\cdot s)
\right]$.
In this approximation, the shape distributions become
\be
\frac1{\sigma_{\rm tot}}{d\sigma_{\rm IR}(e) \over de}
= \sum_N\; |\langle N|W(0)|0\rangle|^2\;
\delta\left( e-e(N)\right)\, ,
\lab{def}
\ee
where $W(0)=\Phi_+(0) \left(\Phi_-(0)\right)^\dagger$ is the total eikonal
phase of
the two quark jets \cite{KS1}, and where the sum is over final states
$|N\rangle$.

As it stands, Eq.\ (\ref{def}) requires renormalization, which we can carry out
with scale $\mu$ serving as a UV
cut-off on soft gluon momenta contributing to $S_{\rm NP}$.
Recalling that $\mu$ appears  as an IR cutoff in the perturbative expansion of
the infrared factor in Eq.\ (\ref{dec}), we may identify $d\sigma_{\rm
IR}(e)/dt$
in \re{def} directly with the infrared shape function for observable $e$.
For example, the shape function
$f(\eps_L,\eps_R)$ corresponds to the particular choice of
$e$ as the pair of light cone variables $(\eps_L,\eps_R)$ defined in Eq.\
\re{MtoMeps}.

To express the shape function as a vacuum expectation value, we introduce an
operator ${\cal E}(\vec n)$ that measures the density of energy flow
in the direction of unit vector $\vec n$ \ci{OreSt}. The operator ${\cal
E}(\vec n)$ is defined
by its action on the final states of $N$ soft gluons with the energy
$E_i$ and momentum $E_i\vec{n}_i$
\be
{\cal E}(\vec n)|N\rangle =
\sum_{i=1}^N\delta(\cos\theta-\cos\theta_i)\; \delta(\varphi-\varphi_i)\;
E_i|N\rangle\, ,
\label{calEdef}
\ee
with $\theta$ and $\phi$ the
spherical angles defining the position of unit vector $\vec n$
with respect to the thrust axis.
It admits a representation in terms of the energy-momentum tensor at
infinity \cite{ST,KOS,T}. Using the definition \re{calEdef},
we find the following operator expression for the shape function
$f(\eps_L,\eps_R)$,
\be
f(\eps_L,\eps_R)=\langle 0 | W^\dagger(0)
\delta\left(\eps_L- \int d \vec n\, w_L(\vec n)\; {\cal E}(\vec n) \right)
\delta\left(\eps_R- \int d \vec n\, w_R(\vec n)\; {\cal E}(\vec n) \right)
W(0) |0\rangle\, ,
\lab{f-double}
\ee
and, as a result of Eq.\ \re{FTdef}, for $f_T(\eps)$,
\be
f_T(\eps)=
\langle 0 | W^\dagger(0)
\delta\left(\eps- \int d \vec n\, w_T(\vec n)\; {\cal E}(\vec n) \right)
W(0) |0\rangle\, .
\lab{ftdef}
\ee
In these expressions, the integration is over unit 3-vector $d\vec n\equiv
d^3 n \,\delta(1-|\vec n|)=d(\cos\theta) d\varphi$, 
and the weight functions, $w_T=w_R+w_L$, are given by%
\footnote{One can show that for the weight function
$w_{-}(\vec n)=1-\cos\theta$ the shape function $f_{-}(\eps)$ coincides with
the inclusive (light-cone) structure function $F(x)$ {\protect\ci{KM}} in
the limit
$1-x=\eps/Q\to 0$.}
\be
w_{R,L}(\vec n)=(1-|\cos\theta|)\,\Theta(\pm \cos\theta)\,,\qquad
w_T(\vec n)=1-|\cos\theta|\,.
\ee
The factorization scale $\mu$ has the meaning of the normalization point of
operators entering \re{ftdef}. We notice that the weight functions suppress
the emissions of gluons collinear to quark jets,
$w(\vec n)\to 0$ as $\cos\theta\to \pm 1$, and that therefore the
shape functions do not suffer from collinear singularities.

The function $f_T(\eps)$ cannot be considered
as a universal distribution for the final states of
${\rm e}^+{\rm e}^-$-annihilation, even in the
two-jet limit, since it reflects the choice of the
weight $w_T$. However, taking its integer moments with respect to $\eps$, \be
\langle \eps^N \rangle\equiv
\int_0^\infty d\eps\; \eps^N\; f_T(\eps,\mu)
=
\int \prod_{i=1}^N\; d{\vec n}_i\, w_T(\vec n_i)\;
{\cal G}(\vec n_1\dots \vec n_N)\, ,
\lab{G}
\ee
we encounter the Green functions $\cal G$ given by multiple
correlators of the energy flow operators ${\cal E}(\vec n)$,
\be
{\cal G}(\vec n_1\dots \vec n_N;\mu)
= \langle 0 | W^\dagger(0)
{\cal E}(\vec n_1)\dots {\cal E}(\vec n_N)   W(0) |0\rangle\, .
\lab{Green}
\ee
These Green functions are defined on a sphere of infinite
radius, at which the energy flow ${\cal E}(\vec n)$ is measured
\ci{ST,KOS,T}.  They depend only
on the $N$ unit vectors $\vec n_i$, in addition to the two light-cone
directions, $p_+$ and $p_-$, of the quarks, represented by the
Wilson lines of Eq. (\ref{def}). In contrast to $f_T(\eps)$, the Green
functions
$\cal G$ are truely universal, depending only on the underlying dynamics
rather than the choice of event shape.

It is interesting to note that a stereographic projection maps
each of the unit vectors $\vec n$ into a point on the
complex plane $\rho=(z,\bar z)$, so that ${\cal G}$ can be
considered as a function on the 2-dimensional $\rho-$plane.
In addition, $\cal G$ is singular only when two or more $\vec n_i$'s
coincide, or equivalently the points $\rho_i$
approach each other on the 2-dimensional plane. These
properties suggest that it possible to relate the Green functions
\re{Green} to $(N+2)-$point correlation functions in some effective
local two-dimensional field theory.

Let us explore further the relation of the functions $\cal G$ to the
thrust, by considering the lowest moments of the thrust distribution
Eq.\ (\ref{dsigdt})
\be
\langle t \rangle
=
{1\over Q}\langle \eps \rangle +\langle t\rangle_\pt
\,,\qquad
\langle t^2 \rangle - \langle t \rangle^2
=
{1\over Q^2}\left[ \langle \eps^2\rangle - \langle \eps \rangle^2 \right]
         + \langle t^2 \rangle_\pt - \langle t \rangle^2_\pt\, .
\lab{variance}
\ee
Here, the moments of $\eps$ are defined through \re{G} and
$\langle ...\rangle_{\pt}$ corresponds to the perturbative contribution.
We observe that perturbative and nonperturbative effects contribute
{\it additively\/} to the moments of the thrust distribution and that the
$1/Q$ correction cancels in the variance of $t$.
The leading nonperturbative corrections enter \re{variance} through the
average values of the powers of $\eps$ in Eq.\ (\ref{avgepsn}),
\bea
\langle \eps \rangle
=\lambda_1 &=& \int d{\vec n}_1 w_T(\vec n_1)\; {\cal G}(\vec n_1)
\nonumber\\
\langle \eps^2 \rangle
- \langle \eps \rangle^2 =-\lambda_2
&=& \int d{\vec n}_1d{\vec n}_2\; w_T(\vec n_1)w_T(\vec n_2)\;
\left [\; {\cal G}(\vec n_1,\vec n_2) -
{\cal G}(\vec n_1){\cal G}(\vec n_2)\; \right ]\, .
\lab{varme}
\eea
We see that the mean value of $\eps$ for the nonperturbative
function $f_T(\eps)$, and hence the $1/Q$ power correction to
$\langle t\rangle$ in Eq.\ (\ref{variance}), is determined by the average
value of
${\cal G}(\vec n)$, weighted by $w_T(\vec n)$. The variance of $\eps$,
however, is related to the ``connected'' part of the
${\cal G}(\vec n_1,\vec n_2)$.

Similar considerations apply to the heavy mass distribution.
Using \re{dsigdrho} we calculate the lowest moments and obtain
\bea
\langle \rho \rangle
&=&
{1\over 2Q}\langle \eps \rangle +\langle \rho\rangle_\pt
\,,
\nonumber\\
\langle \rho^2 \rangle - \langle \rho \rangle^2
&=&  \langle \rho^2 \rangle_\pt - \langle \rho\rangle^2_\pt
+ {1\over 4Q^2}\left[ \langle \eps^2\rangle - \langle \eps \rangle^2 +
\langle \left(\eps_R-\eps_L\right)^2\rangle\, \left\{1+4\int d\rho'\rho'\,
\left({d\sigma^{\rm PT}_J\over d\rho'}\right)^2 \right\}\,
\right]
         \,,
\nonumber\\
\lab{variance1}
\eea
where the nonperturbative scales $\langle\eps^n\rangle\equiv
\langle(\eps_R+\eps_L)^n\rangle$ $(n=1,2)$ are the {\it same\/} as for
the thrust moments and a new scale $\langle (\eps_R-\eps_L)^2\rangle$
is defined as an average with respect to the shape function
$f(\eps_L,\eps_R)$.
We observe that the $1/Q$ correction to the mean value $\langle \rho \rangle$
is half that for $\langle t \rangle$, independent of the
form of the shape functions.  However, because of their
different dependence on $\eps_L$ and $\eps_R$, starting from the second moment
nonperturbative corrections to the thrust and heavy
jet mass are not simply related,
unless one chooses a particular ansatz for the underlying shape function
$f(\eps_L,\eps_R)$.

The functions ${\cal G}$ defined in Eq.\ \re{Green} measure the energy flow in
the final state, and it is not surprising therefore that the energy-energy
correlation (EEC),
\be
{\rm EEC}(\chi)={1\over Q^2} \sum_{i,j}E_iE_j\;
\delta(\cos\chi-\cos\theta_{ij})\, ,
\lab{eecdef}
\ee
has a simple interpretation in terms of these
correlation functions.
The leading $1/Q$ nonperturbative
contribution to the EEC is associated with soft gluons
propagating in the direction defined
by the angle $\chi$, and it may thus be written as the expectation value of
a single ${\cal G}(\vec n)$.  Away from the jet directions, the soft gluon
radiation that contributes to the EEC factorizes from the jets, and
the EEC may be written as
the usual expectation value for soft radiation
from a product of eikonal lines,
\be
\langle {\rm EEC}(\chi)\rangle_{_{\rm NP}}
={1\over Q}\langle 0|W^\dagger(0)\; {\cal E}(\vec n)\;
W(0)\; |0\rangle={1\over Q}{\cal G}(\chi)\, ,
\lab{eeccalE}
\ee
where $\chi$ is the angle between $\vec n$ and the two-jet axis.
We emphasize that this result holds only for $\chi$
fixed and away the directions $\chi= 0,\pi$,
where $Q\sin\chi$ is a new, small parameter.
Comparing Eq.\ \re{eeccalE} and
Eqs.\ \re{varme} and \re{variance} we observe a relation between
the leading $1/Q$ corrections to ${\rm EEC}(\chi)$ and mean value of the
thrust. This relation depends, however, on the angular form of the function
${\cal G}(\chi)$.

We can now relate our formalism to previous recent work on power corrections.
Consider first the simplest ansatz for the ${\cal G}$'s, in which
all higher-order connected parts of the Green functions $\cal G$ are
neglected,
\be
{\cal G}(\vec n_1,\dots \vec n_N)=\prod_{i=1}^N{\cal G}(\vec n_i)\, .
\lab{ansatz}
\ee
According to Eqs.\ \re{G} and \re{f-double}, this is equivalent to
replacing the
shape functions $f_T(\eps)$ and $f(\eps_L,\eps_R)$ by delta functions,
\be
f_T(\eps)=\delta(\eps-\lambda_1)\,,
\qquad
f(\eps_L,\eps_R)=
\delta\left(\eps_L-\frac12\lambda_1\right)
\delta\left(\eps_R-\frac12\lambda_1\right)\, .
\lab{ftdelta}
\ee
In this case, the variances of the nonperturbative distributions in
$\eps$ vanish, and substitution of \re{ftdelta} into \re{dsigdt}
yields a simple shift in the perturbative thrust distribution, as in
Eq.\ \re{shift}. The same happens for the heavy mass distribution,
but with half the shift in $\rho$.

These event shape power corrections can be further related to the energy-energy
correlation, if a form is given for the function ${\cal G}(\vec n)$
itself.
If we assume, for example, that the angular dependence of
${\cal G}$ is identical to that of single-gluon emission diagrams,
${\cal G}(\vec n) \sim 1/\sin^3\chi$, Eq.\ \re{varme} implies that
\be
{\cal G}(\chi)={\lambda_1\over 4\pi\, \sin^3\chi}\, ,
\lab{Goneg}
\ee
which relates the $1/Q$ power correction of the thrust
to the EEC.  In the model of \ci{Milano}, the
parameter $\lambda_1$ is related to a nonperturbative,
``universal" coupling constant $\alpha_s^{\rm NP}$,
although our analysis does not require such an interpretation.
Comparison of the form Eq.\ \re{Goneg} to data will
help gauge the size of corrections associated
with the variance and higher moments of the Green functions $\cal G$.

There is evidence that corrections beyond
Eq.\ \re{ftdelta} are necessary.  For the thrust distribution, the
simple shift of Eq.\ \re{shift}
improves agreement
with the data \cite{DW}, relative to resummed perturbation
theory, for $t >
\Lambda_{\rm QCD}/Q$, but for smaller values of
$t\sim \Lambda_{\rm QCD}/Q$
 higher-order connected correlations
evidently become important.
The variance, for example, includes
information on branching of final state gluons in all possible relative
configurations, including opposite hemispheres \cite{NS}. Its contribution to
(\ref{ftdelta}) smears the $\delta-$function distribution
and gives rise to a ``nonfactorized'' correction to the shape function,
$\Delta f(\eps_L,\eps_R)$ in Eq.(\ref{ffact}). As a simple
ansatz for the shape functions that models these effects, one may consider
the following
expressions \ci{Moriond98}
\bea
f(\eps_L,\eps_R)&=& \left(\frac{\eps_L\eps_R}{\Lambda^2}\right)^{\alpha-1}
\exp\left(-\frac{(\eps_L+\eps_R)^2}{\Lambda^2}\right)
\frac{2\Gamma(2\alpha)}{\Lambda^2\Gamma^3(\alpha)}\,,
\nonumber
\\
f_T(\eps)&=&\left(\frac{\eps}{\Lambda}\right)^{2\alpha-1}
\exp\left(-\frac{\eps^2}{\Lambda^2}\right)
\frac{2}{\Lambda\Gamma(\alpha)}\, ,
\eea
with $\alpha$ and $\Lambda$ free parameters. Their values can be
fixed using (\ref{varme}) as
\be
\Lambda\frac{\Gamma(\alpha+\frac12)}{\Gamma(\alpha)}=\lambda_1\,,\qquad
\Lambda^2\alpha=\lambda_1^2-\lambda_2\,.
\ee
A fit of this kind for $f_T$, using data at a single value of
$Q$ -- the Z mass -- induces a good description of the data for the thrust
distribution for energies $14~{\rm GeV} \le Q \le 161~{\rm GeV}$,
over the full range $0<t<1/3$ as shown in Fig.~1 \ci{Moriond98,HS}. 
Further analysis of event shape energy
dependence should make it possible to estimate the underlying
energy flow functions ${\cal G}(\vec n_i)$.

\begin{figure}[ht]
\centerline{\epsfysize=12cm\epsfxsize=14cm
\epsffile{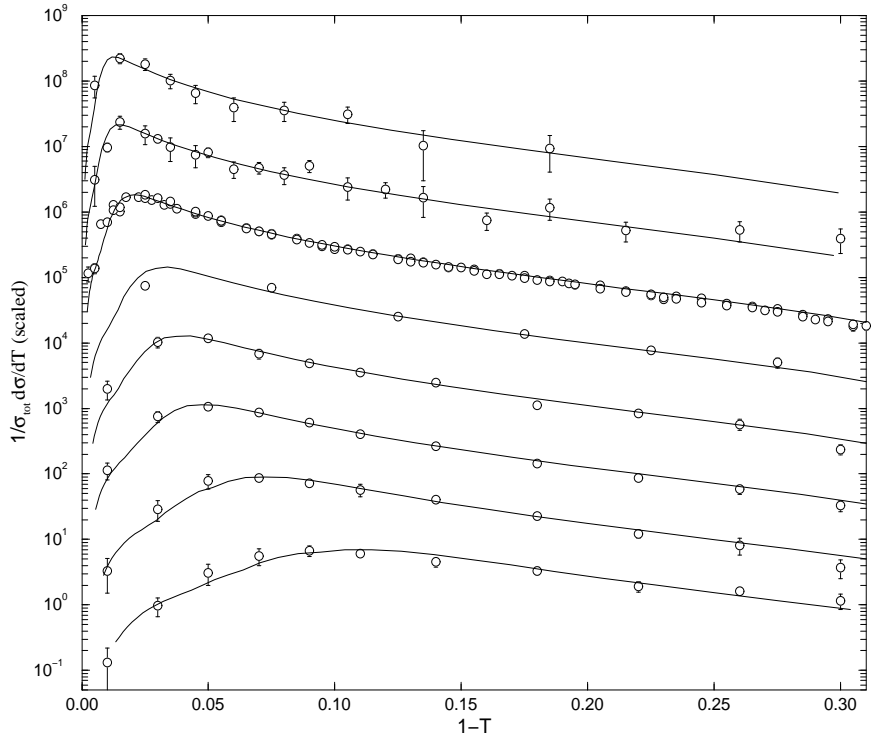}}
\vspace*{-7mm}
\caption{The comparison of the data with the QCD prediction for the 
thrust distribution at different energies (from bottom to top):
$Q/{\rm GeV}=14,22,35,44,55,91,133,161,$ based on the
shape function. The detailed description of the plot can be found in
\ci{Moriond98}.}
\end{figure}

\section{Summary}

In this paper we have studied the power corrections to the
differential thrust, $t=1-T$, and heavy mass, $\rho$, distributions
in ${\rm e}^+{\rm e}^-$ annihilation close to the two-jet limit.
In addressing this problem, we did not aim to
justify a particular QCD-inspired phenomenological model,
but rather to formulate a framework with which to study
the relationship between perturbative and nonperturbative effects
in high-energy final states. We have
seen that, despite the fact that the thrust and heavy jet mass
are not inclusive quantities, the leading nonperturbative corrections
to their differential distributions can be factorized into the
perturbative and nonperturbative functions,
in much the same way as for inclusive cross sections.
We identified nonperturbative infrared shape functions
that organize all leading power corrections, $1/(tQ)^n$ and $1/(\rho Q)^n$.
Although not universal themselves, these shape functions can be derived
from universal matrix elements that describe energy flow.
We anticipate that it will be possible to extend these
considerations to a wide class of infrared safe event shapes and
hard-scattering
processes.

\subsection*{Acknowledgements}

We are grateful to Hasko Stenzel for useful correspondence.
This was supported in part by the EU Fourth Framework Programme ``Training
and Mobility of Researchers'', Network ``Quantum Chromodynamics and the Deep
Structure of Elementary Particles'', contract FMRX--CT98--0194 (DG 12 -- MIHT)
and by the National Science Foundation, grant PHY9722101.

\end{document}